\documentclass[paper,prd,reprint,preprintnumbers,nofootinbib,notitlepage,showpacs,showkeys]{revtex4-1}
\usepackage{latexsym,graphicx,amssymb,amsmath,mathrsfs} 

\newcommand{\be}{\begin{equation}}    
\newcommand{\ee}{\end{equation}}    
\newcommand{\bea}{\begin{eqnarray}}    
\newcommand{\eea}{\end{eqnarray}}    
    
\newcommand{\rt}[1]{{}}    
\renewcommand{\div}{\,\textrm{div}\,}

\begin{document}
\allowdisplaybreaks

\title{Renormalized $O(N)$ model at next-to-leading order of the $1/N$ expansion: Effects of the Landau pole}

\author{G. Fej\H{o}s}
\email{fejos@riken.jp}
\affiliation{Theoretical Research Division, Nishina Center, RIKEN, Wako 351-0198, Japan}

\author{A. Patk{\'o}s}
\email{patkos@galaxy.elte.hu}
\affiliation{Department of Atomic Physics, E{\"o}tv{\"o}s University, H-1117 Budapest, Hungary}

\author{Zs. Sz{\'e}p}
\email{szepzs@achilles.elte.hu}
\affiliation{MTA-ELTE Statistical and Biological Physics Research Group, H-1117 Budapest, Hungary}

\preprint{RIKEN-QHP-148}

\begin{abstract}
Apparently convergent contributions of resummed perturbative series at the next-to-leading order of the $1/N$ expansion in the $O(N)$ model are reanalyzed in terms of renormalizability. Compared to our earlier article [G. Fej\H{o}s {\it et al.}, Phys.\ Rev.\ D {\bf 80}, 025015 (2009)], an additional subtraction is performed. We show numerically that this is indispensable for diminishing the cutoff sensitivity of some integrals below the scale of the Landau pole. Following the method of our earlier article, an improved counterterm Lagrangian is constructed in the two-particle irreducible formalism, with and without the use of an auxiliary field formulation. 
\end{abstract}

\pacs{11.10.Gh, 12.38.Cy}
\keywords{Renormalization; large-$N$ approximation; 2PI formalism}

\maketitle

\section{Motivation}

The study of the renormalizability of the $O(N)$ model at next-to-leading order (NLO) of the $1/N$ expansion started fairly early \cite{Root:1974zr}. Recently an attempt to renormalize the effective potential at this level of the expansion was reported in \cite{Andersen:2008qk}, where the pressure of the pion-sigma gas at finite temperature was calculated with its help. In that article the auxiliary field formulation of the model was used and the renormalization was achieved only at a given point of the auxiliary field configurations obtained by exploiting its saddle point equation explicitly. In \cite{Fejos:2009dm} we presented an explicit construction of the NLO counterterms with and without the introduction of the auxiliary field and found that the model is actually renormalizable for arbitrary values of the field expectation values. The zero temperature divergence structure of the dynamical equations derived from the two-particle irreducible (2PI) effective action was investigated. This form of the action depends independently on the fields and the corresponding propagators. A strict $1/N$ expansion of the pion propagator has been performed which changes the self-consistent nature of the 2PI-$1/N$ approximation into a hierarchical structure. An important feature of this procedure is that, at the NLO level of the approximation scheme, the propagators are given explicitly in terms of the leading order (LO) expressions and this makes the analysis of the asymptotic behavior of various loop integrals more transparent than of the 2PI-$1/N$ approximation whose renormalization was treated in \cite{Cooper:2005vw,Berges:2005hc}.

The appearance of a tachyonic pole (the Landau singularity) has been observed to be a fundamental feature of the large-$N$ approximation already in the pioneering publications \cite{Schnitzer:1974ji,Coleman:1974jh,Root:1974zr}. Thorough studies led to the understanding of the effective nature of the renormalized $O(N)$ model in which the cutoff cannot be sent to infinity. The $1/N$ expansion turned out to be a valuable tool for studying phenomena dominated by momentum scales well below the cutoff, which is chosen to be substantially smaller than the scale of the Landau singularity \cite{Abbott:1975bn,Bardeen:1983st,Nunes:1993bk}.

In our previous work \cite{Fejos:2009dm} we analyzed the integrals exploiting the asymptotic behavior of their integrands for infinitely large momenta, which is a customary procedure in the perturbative analysis of divergences. However, this approach turns out to be somewhat ambiguous and needs to be corrected, since due to the explicit presence of the Landau pole in these integrals, actually it is not possible to send any momenta to infinity. In particular, this limitation restricts the range of cutoff values applied for the regularization of divergent integrals. In this context, the meaning of renormalization is actually to achieve a practical cutoff insensitivity below the scale of the Landau pole, similarly to the cases discussed in \cite{Reinosa:2011ut, Marko:2012wc, Marko:2013lxa}. The Landau singularity affects both the structure and the explicit expression of the counterterms. As it will become clearer in the next section, taking it into account becomes necessary because in the counterterm functional we omitted to include contributions of the form 
\be
\int d^4p\frac{1}{(p^2-M_0^2)^2\ln^2(p/\Lambda_p)}\sim \int^\Lambda \frac{d p}{p}\, \frac{1}{\ln^2(p/\Lambda_p)},
\label{Eq:new-int}
\ee
where on the right-hand side, valid for large momenta, a cutoff regularization was used. In this integral one cannot neglect the presence of $\Lambda_p,$ which is the value of the Landau pole and conclude, as in \cite{Fejos:2009dm, Cooper:2005vw}, that the integral behaves as $1/\ln\Lambda.$ Actually, the integral diverges as $\Lambda\to\Lambda_p$ and the right question to be asked is in which part of the region $\Lambda<\Lambda_p$ the Landau pole influences the cutoff dependence of the integral. Of course, if one asked what the cutoff dependence of the integral beyond the scale of the Landau pole is, and in fact this is what we did in \cite{Fejos:2009dm}, then one would conclude that this behavior is $1/\ln\Lambda$. But even though in a mathematical sense correct, this would be an answer in a rather unphysical situation. 

The most important goal of this paper is to complete our divergence analysis presented in \cite{Fejos:2009dm} by investigating whether the subtraction throughout the calculation of a previously omitted integral, similar to that in \eqref{Eq:new-int}, can be done in a way consistent with the requirements of the counterterm renormalization applied to the resummed perturbative series provided by the $1/N$ expansion. In the next section we shall analyze some of the relevant integrals and study numerically the changes observed in their cutoff dependence after performing appropriate subtractions. For these integrals we shall use cutoff regularization, with a cutoff chosen below the scale of the Landau pole. Although in this case all the integrals are strictly finite, we shall still call an integral "divergent'' ("convergent'') if for increasing momentum $p<\Lambda <\Lambda_p$, its integrand decreases slower (faster) than $1/p^4$ (up to logs). In Sec.~III we summarize those results of Ref.~\cite{Fejos:2009dm} which are directly needed for our present purpose, but we also provide guidance to the relevant parts of the original paper. The renormalization procedure presented in Sec.~\ref{sec:ren} requires appropriate subtractions to be imposed on any integral called divergent in the above sense in order to diminish its cutoff sensitivity for increasing $\Lambda,$ already below the Landau pole. Only by sending the cutoff to infinity in some finite integrals entering the integrand of divergent integrals, we could obtain analytical expressions for the divergent part. We shall discuss in the concluding Sec.~\ref{sec:concl} the criterion a consistent cutoff regularization should satisfy and the calculational difficulties posed by a regularization, in which all propagator momenta in an integral are kept below the value of the cutoff. Also, the possibility of oversubtractions, further diminishing the sensitivity of specifically chosen $n$-point functions to the presence of the Landau pole, is shortly assessed.

\section{Subtraction method and cutoff dependence of the integrals}
The NLO equations of the $1$- and $2$-point functions can be written in terms of an effective momentum-dependent coupling ($\lambda$ is the renormalized coupling)
\be
\label{Eq:eff_cpl}
\lambda(p)=\frac{\lambda}{1-\frac{\lambda}{6}I_\pi^F(p)},
\ee
which reflects that the LO solution of the $1/N$ expansion effectively resums an infinite series of pion bubbles. The analysis presented in \cite{Fejos:2009dm} shows that the divergences in these equations are momentum independent and given by local integrals which are elements of a class of integrals characterized by integers $j,k$ satisfying $j\geq 1, k\geq 0$:
\be
\label{Eq:I_def}
I^{j,k}=(-i)^{j-1} \int_p D_\pi^j(p) \lambda^k(p),
\ee
where we used the shorthand notation $\int_p\equiv \int\frac{d^4p}{(2\pi)^4}$ and $D_\pi(p)=i/(p^2-M^2)$ is the tree-level pion propagator. The finite part of the bubble integral $I_\pi(p)=-i\int_k D_\pi(k)D_\pi(p+k)$ is defined as $I_\pi^F(p)=-i\int_k \left[D_\pi(k)D_\pi(p+k)-G_0^2(k)\right]$, where $G_0(k)=i/(p^2-M_0^2)$ is an auxiliary propagator in which $M_0$ plays the role of the renormalization scale.

One has to investigate carefully the asymptotic behavior of the integrands in \eqref{Eq:I_def} in order to find out which of them needs subtraction and to assess those appropriate subtractions which efficiently diminish the sensitivity of the integrals with respect to an increasing cutoff, but still below the (Landau) pole of the effective coupling \eqref{Eq:eff_cpl}. For us, the most important element of the set $I^{j,k}$ is $I^{2,2}$ because it was left unsubtracted in \cite{Fejos:2009dm} due to a formal logarithmic power counting, which was not careful enough, as explained in the previous section. 

Formal power counting based on the asymptotic behavior of the integrands in \eqref{Eq:I_def} for $p\to\infty$ suggests that if $j>3$ ($j\leq 2$), then $I^{j,k}$ should be considered convergent (divergent). This is certainly true for $k=0,$ a case in which $I^{1,0}$ is the tadpole integral and $I^{2,0}$ is the bubble integral at vanishing external momentum. Their respective finite parts can be defined through the minimal subtraction renormalization scheme used in \cite{Fejos:2009dm}. To see what is to be subtracted we expand $D_\pi(p)$ around $G_0(p)$ 
\be
\label{Eq:prop_expand}
D_\pi=G_0-i(M^2-M_0^2)G_0^2-(M^2-M_0^2)^2 D_\pi G_0^2.
\ee
Using the quadratically and logarithmically divergent integrals introduced in \cite{Fejos:2009dm},
\be
\label{Eq:Td2_0}
T_d^{(2)}=\int_k G_0(k), \qquad T_d^{(0)}=-i\int_k G^2_0(k),
\ee
and the notation
\be
\label{Eq:td}
t_d(M^2)=T_d^{(2)}+(M^2-M_0^2)T_d^{(0)},
\ee
the subtraction implied by \eqref{Eq:prop_expand} gives the following finite integrals:
\begin{subequations}
\label{Eq:subt1}
\bea
\label{Eq:subt1_T}
T_\pi^F&:=&I^{1,0}_F=\int_p D_\pi(p) - t_d(M^2), \quad\\
I_\pi^F(k=0)&:=&I^{2,0}_F=-i\int_p D_\pi^2(p)-T_d^{(0)}.
\label{Eq:subt1_I}
\eea
\end{subequations}

As already discussed in the previous section, for $k\ne0$ one has to perform a more careful analysis because the integrands explicitly display the Landau pole at a value of the momentum which depends on the coupling approximately as $\Lambda_p\approx M_0\exp\big(1+48\pi^2/\lambda\big).$ In this case, in addition to \eqref{Eq:prop_expand}, one needs also to expand \footnote{The expansion induces a new pole related to $M_0$, but unless there is a huge difference between the masses, the two poles are very close to each other. The location of the new pole is lower (higher) than the original for $M_0>M$ ($M_0<M$). Since the pion mass $M$ vanishes in the chiral limit, we choose $M_0>M,$ in which case, for a given $\lambda$, the singularity of a subtracted integral is basically determined by $M_0.$}  $\lambda(p)$ around $\lambda_0(p)=\lambda/(1-\lambda I_0^F(p)/6)$, 
\bea
\lambda(p)&=&\lambda_0(p)+\frac{1}{6}\lambda_0(p) \lambda(p)\big[I_\pi^F(p)-I_0^F(p)\big],
\label{Eq:eff_cpl_expand}
\eea
where $I_0^F(p)$ is obtained by replacing $D_\pi$ with $G_0$ in the definition of $I_\pi^F(p)$ given below \eqref{Eq:I_def}. To analyze the divergence of some integrals, {\it e.g.} of $I^{1,1},$ one needs to know explicitly the expansion of the difference $I_\pi^F(p)-I_0^F(p)$ for large momenta, which is given later in \eqref{Eq:I_expand}. However, for the integrals $I^{2,1}$ and $I^{2,2}$ presented in this section to illustrate the effect of the Landau pole, we only need to know that this difference is ${\cal O}(1/p^2).$ Using this fact and the leading order terms in the expansions \eqref{Eq:prop_expand} and \eqref{Eq:eff_cpl_expand}, one finds for the respective minimally subtracted finite parts:
\begin{subequations}
\label{Eq:subt2}
\bea
I^{2,1}_F&=&I^{2,1} +i\int_p G_0^2(k)\lambda_0(k) = I^{2,1}-\lambda\, t_a,\\
I^{2,2}_F&=&I^{2,2}-\lambda^2 T_a^{(0)},
\eea
\end{subequations}
where we introduced the notation
\be
\label{Eq:ta}
t_a=T_a^{(0)}-\frac{\lambda}{6}T_a^{(I)}
\ee
for the combination of the two integrals in terms of which the subtractions are defined. The integral
\bea
\label{Eq:TaI}
T_a^{(I)}=-i\int_k G^2_a(k)I_0^F(k),
\eea
was already defined in \cite{Fejos:2009dm} with the auxiliary propagator $G_a(k)=i/[(k^2-M_0^2)(1-\lambda I_0^F(k)/6)],$ while 
\bea
\label{Eq:Ta0}
T_a^{(0)}=-i\int_k G^2_a(k), 
\eea
introduced in analogy with $T_d^{(0)},$ is a new integral which was not subtracted in our previous analysis. The last divergent integral we shall use from \cite{Fejos:2009dm} is
\be
\label{Eq:Ta2}
T_a^{(2)}=\int_k G_a(k).
\ee

\begin{figure}[!t]
\begin{center}
\includegraphics[width=0.49\textwidth,angle=0]{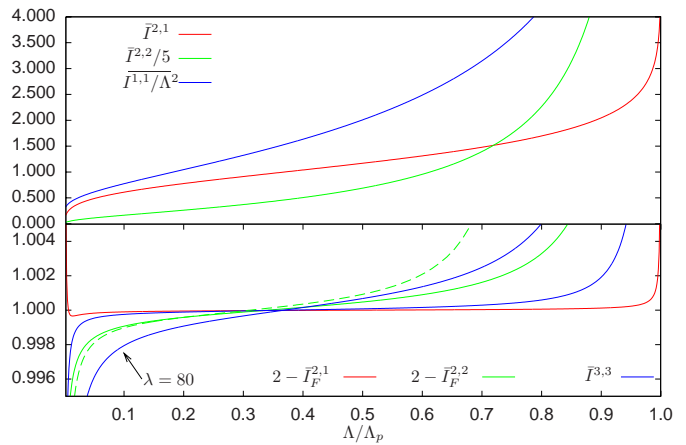}
\caption{The cutoff dependence of divergent integrals (upper part) and subtracted or convergent integrals (lower part) computed by taking $\Lambda\to\infty$ in the finite bubble integrals $I_\pi^F(p)$ and $I_0^F(p)$, with the exception of the dashed curve corresponding to $j=k=2,$ for which the cutoff regularization given in \eqref{Eq:2theta_reg} is applied. The bar on $I^{j,k}$ indicates that the integrals are scaled by the value taken at the inflection point (inflection of $I^{1,1}/\Lambda^2$ when $j=k=1$) and, for the sake of the presentation, also by an additional factor in the case of $j=k=2.$ We set $M=1,$ $M_0/M=2,$ and used $\lambda=65$ for the coupling, except where indicated.}
\label{Fig:Inm}
\end{center}
\end{figure}

In the remaining of this section we study numerically the cutoff dependence of the simplest integrals of the $I^{j,k}$ set. In the upper part of Fig.~\ref{Fig:Inm} we see that just like $I^{2,1},$ the integral $I^{2,2}$ does not show any practical cutoff insensitivity below the Landau pole $\Lambda_p,$ therefore similarly to $I^{2,1}$, it has to be considered divergent and an appropriate subtraction has to be applied to it. This shows explicitly that from a physical point of view the renormalization of the $O(N)$ model at next-to-leading order in the $1/N$ expansion presented in \cite{Fejos:2009dm} is incomplete: the subtraction of $T_a^{(0)}$ is needed and its effect has to be taken into account in the divergence analysis of other integrals, as well.

In the lower part of Fig.~\ref{Fig:Inm} we see that after applying the subtractions introduced in (\ref{Eq:subt2}) a practical cutoff insensitivity is reached for $\Lambda<\Lambda_p.$ We can roughly say that the region of apparent convergence is limited from above by the inflection point of $I_F^{j,k}(\Lambda)$ for $j=2$ and of $I^{j,k}(\Lambda)$ for $j>2,$ and that as the cutoff is increased above this point the $\Lambda$ dependence of the subtracted integrals becomes clearly dominated by the Landau pole. Note that we divide the integrals with the value taken at the corresponding inflection point, in order to be able to show their cutoff dependence in a single plot. As expected, for a given $\lambda$ and $j,$ the plateau-like behavior of the subtracted integrals is more visible at smaller values of $k$ (compare at $j=2$ the curves for $k=1$ and $k=2$). Also, decreasing $\lambda$ makes the $\Lambda$ dependence more flat. As visible in Fig.~\ref{Fig:Inm} for the $j=k=2$ case, below the inflection point of $I^{2,2}(\Lambda)$ it is practically irrelevant if the finite bubble integrals $I_\pi^F$ and $I_0^F$ are computed with the actual finite cutoff or with an infinite one. This is an important observation because the divergence analysis of the next section needs some explicit expressions and we could obtain them only in the latter case. 

For $j>2$, a practical cutoff insensitivity is expected at high enough values of the cutoff, but we would like to stress that this only happens if the value of the coupling, which governs the location of the Landau pole, is not too big. For example, by increasing the coupling from $\lambda=65$ to $\lambda=80,$ we see in Fig.~\ref{Fig:Inm} that the slope of $I^{3,3}$ increases and that the integral becomes sensitive to the presence of the Landau pole at lower values of the cutoff. Therefore, even though each individual term of the series obtained with an expansion of $I^{3,3}$ in powers of $\lambda$ is finite, one may even try to treat these integrals as divergent and define their finite parts with appropriate subtractions. One can imagine doing this procedure gradually, that is starting with convergent integrals having the smallest $j$ value. The question whether this kind of oversubtraction can be realized without any restriction on the choice of the renormalized couplings and for general values of the backgrounds ($v$ of the $\sigma$ field and $\hat\alpha$ of the auxiliary field introduced below) is beyond the scope of the present investigation. It might be of physical interest, therefore we return to this point in the concluding Sec.~\ref{sec:concl}. Throughout the paper we shall assume that the coupling is not very large, meaning that the position of the Landau pole in momentum space is much larger than the physically relevant scales. In this case no subtraction has to be applied to the integrals $I^{j>2,k}.$
\bigskip
\section{The model in the auxiliary field formulation \label{sec:form}}

The next-to-leading order 2PI effective potential of the $O(N)$ model in the $1/N$ expansion (denoted by $\Gamma[\hat{\alpha},v,G_{\pi},{\cal G}]$ in \cite{Fejos:2009dm}) has the following expression:
\begin{widetext}
\bea
\label{Eq:V_2PI}
V[\hat{\alpha},v,G_{\pi},{\cal G}]&=&\frac{N}{2}M^2v^2+\frac{3N}{2\lambda}\hat\alpha^2-\frac{i}{2}\int_k\big[(N-1)\big(\ln G_\pi^{-1}(k)+D_\pi^{-1}(k)G_\pi(k)\big)
+{\textrm{Tr}}\ln{\cal G}^{-1}(k)+{\textrm{Tr}}({\cal D}^{-1}(k){\cal G}(k))
\big]\nonumber\\
&+&i\frac{\lambda}{12}\int_k\int_pG_{\alpha\alpha}(k)G_\pi(p)G_\pi(p+k)+\Delta V[\hat{\alpha},v,G_{\pi},{\cal G}].
\eea
\end{widetext}
We refer to \cite{Fejos:2009dm} (see also \cite{Mihaila:2000sr,Aarts:2002dj}) for details concerning its derivation with the usual rules of the 2PI formalism from the Lagrangian of the model obtained after the elimination of the quartic interaction term through a Hubbard-Stratonovich transformation. Here it is sufficient to know that $v$ is the vacuum expectation value of the dynamical field pointing in the $\sigma$ direction, $m^2$ and $\lambda$ represent the renormalized mass and coupling constant, $\hat\alpha$ is the (rescaled) auxiliary field, and that we use the shorthand notation 
\be 
M^2=m^2-i\hat\alpha.
\ee 
$D_\pi$ is the tree-level pion propagator introduced below \eqref{Eq:I_def}, while $\cal D$ and $\cal G$ are the tree-level and the full $2\times 2$ symmetric propagator matrices in the coupled $\sigma-\alpha$ sector, respectively. The matrix elements of the inverse ${\cal D}^{-1}$ are $(D^{-1})_{\sigma\sigma}(p)=D^{-1}_\pi(p),$ $(D^{-1})_{\alpha\alpha}=i,$ and $(D^{-1})_{\alpha\sigma}=v(\lambda/3)^{1/2}.$ 

The components of ${\cal G}^{-1}$ are obtained from the stationarity condition $\delta V/\delta {\cal G}=0$. Inverting the matrix ${\cal G}^{-1}$ at LO [that is in the case without the last integral of \eqref{Eq:V_2PI}] the components of ${\cal G}^{(0)}$ are given explicitly in Eq.~(15) of \cite{Fejos:2009dm}. The expressions of the $\sigma\sigma,$ $\alpha\alpha,$ and $\alpha\sigma$ matrix elements of the LO propagator matrix in the $\alpha-\sigma$ sector can be conveniently rewritten for the next discussion in the following form:
\begin{subequations}
\bea
\label{Eq:Gss}
G_{\sigma\sigma}^{(0)}(p)&=&D_\pi(p)-i\frac{v^2}{3}\lambda(p)D_\pi(p)G_{\sigma\sigma}^{(0)}(p),\\
\label{Eq:Gaa}
\lambda G_{\alpha\alpha}^{(0)}(p)&=&-i\lambda(p)D_\pi^{-1}(p) G_{\sigma\sigma}^{(0)}(p),\\
\label{Eq:Gas}
\sqrt{\frac{\lambda}{3}}G^{(0)}_{\alpha\sigma}(k) &=& i\frac{v}{3}\lambda(k) G_{\sigma\sigma}^{(0)}(k).
\eea
\end{subequations}
We shall call $G_{\sigma\sigma}^{(0)}(p)$ the LO sigma propagator.

The expressions \eqref{Eq:Gss} and \eqref{Eq:Gaa} prove useful if one wants to see how the integrals $I^{j,k}$ introduced in \eqref{Eq:I_def} are generated. With $G_{\alpha\alpha}^{(0)}$ taken from \eqref{Eq:Gaa} and with $G_{\sigma\sigma}^{(0)}(p)$ used iteratively from \eqref{Eq:Gss} one sees that $I^{j,k}$ with $j=k\ge1$ appears through the integral 
\be
\label{Eq:int_Gaa}
i\lambda \int_p G_{\alpha\alpha}^{(0)}(p) D_\pi(p),
\ee
which appears in the NLO pion self-energy, as we shall see shortly. 

Integrals with $j=k+1,k\ge1$ will be shown to emerge in the functional derivative of the effective action with respect to $\hat\alpha$ through the integral
\be
\label{Eq:int_of_diff}
\frac{i}{2}\int_p\big( G_{\alpha\alpha}^{(0)}(p)-D_\pi(p)\big)=\frac{v^2}{6}\int_p \lambda(p)D_\pi(p) G_{\sigma\sigma}^{(0)}(p),
\ee
again with the iteration of $G_{\sigma\sigma}^{(0)}(p)$ from \eqref{Eq:Gss}.

The counterterm functional $\Delta V$ in \eqref{Eq:V_2PI} contains all the counterterms we need to determine in order to renormalize the effective potential, its functional derivatives with respect to $v$ and $\hat\alpha,$ and the propagators. It is convenient to split $\Delta V$ into several pieces:
\bea
\label{Eq:cts}
\Delta V = \Delta V_{\alpha}^N+\Delta V_{G_{\alpha\alpha}}^{0}+\Delta V_{G_\pi}^{0}+\Delta V_{v}^{0}+\Delta V_\alpha^{0},
\eea
each corresponding to the renormalization of a specific functional derivative of $V$, denoted by the corresponding subscript. The upper indices distinguish between terms corresponding to different orders of the large-$N$ hierarchy. Our task in what follows is to analyze the divergences of the functional derivatives of (\ref{Eq:V_2PI}) in order to find the appropriate choice of terms in \eqref{Eq:cts} which renders these quantities finite.

The leading order renormalization remains unchanged compared to \cite{Fejos:2009dm}, thus we just recall that
\bea
\Delta V_{\alpha}^{N}&=&i\hat{\alpha}\frac{N}{2}t_d(m^2)+\hat\alpha^2\frac{N}{4}T_d^{(0)},\ \ \\
\Delta V_{G_{\alpha\alpha}}^{0}&=&\frac{\lambda}{12}T_d^{(0)}\int_k G_{\alpha\alpha}^{(0)}(k),
\eea
with the function $t_d(x)$ introduced in \eqref{Eq:td} and $T_d^{(0)}$ given in \eqref{Eq:Td2_0}.

At NLO we have to deal with three quantities: the pion propagator and the derivatives $\delta V/\delta v$ and $\delta V/\delta\hat\alpha.$ We expect that, although renormalizability should not work for arbitrary propagators, as the proof of perturbative renormalizability strongly relies on their asymptotic behavior, there must be no restriction concerning the value of the background fields $v$ and $\hat\alpha$. For this reason, we shall analyze the corresponding derivatives of the effective potential instead of the field equation for $v$ and the saddle point equation for $\hat\alpha$ which arise by equating the respective derivatives to zero. 

We would like to note that, since we employ, as did also in \cite{Fejos:2009dm}, a strict $1/N$ expansion in the pion propagator, the divergence analysis at NLO could be equally well performed within the 1PI formalism. In particular, the counterterms determined here are not the ones rendering finite the effective potential of the 2PI formalism truncated at two-loop level and the self-consistent pion propagator derived from it, but rather they should be understood as approximating those. For the determination of the full 2PI counterterms using self-consistent equations the reader should consult Ref.~\cite{Berges:2005hc}. Our use of the 2PI formalism is motivated by the fact that the highly nontrivial resummation of infinite classes of diagrams in the effective potential can be rather compactly formulated by combining it with the auxiliary field formulation of the model. This is because at the NLO level of the $1/N$ expansion only the contribution of a single two-loop 2PI integral has to be taken into account in \eqref{Eq:cts}. In \cite{Fejos:2009dm}, the fact that field and propagator are independent variables of a common 2PI effective potential was used as a tool for organizing our analysis. It facilitates tracking the influence of a counterterm piece determined from a certain derivative of the effective potential on the renormalization of another derivative. The introduction of the auxiliary field explicitly provided guidance for the renormalization of the pion self-energy also in the case when the auxiliary field was not used, as it indicated that the right strategy to follow is to independently renormalize the momentum-dependent and momentum-independent parts of the self-energy.

\section{Next-to-leading order renormalization \label{sec:ren}}
Before presenting the detailed renormalization steps leading to the completion of the list of counterterms determined in \cite{Fejos:2009dm}, we point out the changes in the final result, as compared to our previous analysis:
\begin{itemize}
\item 
The divergence of the pion propagator equation changes in two ways. First, the expression of $\tilde{T}_{\div}(M^2)$ given in Eq.~(21) of \cite{Fejos:2009dm} changes such that the "double scoop'' integral $\left(\int_k G_\pi(k)\right)^2$ and also a term proportional to $v^2 \int_k G_\pi(k)$ are induced in the $\Delta V_\pi^0$ piece of the counterterm functional. Second, there also appears an additional divergence proportional to $v^2$ in the integral \eqref{Eq:int_Gaa} [Eq.~(20) of \cite{Fejos:2009dm}]. This new term modifies the divergence of $\delta V/\delta v$ and induces a new counterterm proportional to $v^4$ in $\Delta V_v^0.$ Interestingly, these new terms combine in the counterterm functional into a term proportional to $\left(v^2+\int_k G_\pi(k)\right)^2$, which, however, has no renormalized counterpart in the auxiliary field formulation of the model given in \eqref{Eq:V_2PI}.
\item 
The divergence analysis of the derivative $\delta V/\delta\hat\alpha$ also changes because the divergences of the integrals $J(M^2)$ and $\tilde{J}(M^2)$ introduced in Eq.~(29) of \cite{Fejos:2009dm} have to be reanalyzed, as terms proportional to $T_a^{(0)}$ were previously not included.
\end{itemize}

Now we go into the details. 

\subsection{The NLO pion propagator equation.} 

This equation reads as
\bea
\label{Eq:pion}
iG_{\pi}^{-1}(k)&=&iD_{\pi}^{-1}(k)-i\frac{\lambda}{3N}\int_p G_{\alpha\alpha}^{(0)}(p) D_\pi(p)\nonumber\\
&-&i\frac{\lambda}{3N}\int_p G_{\alpha\alpha}^{(0)}(p)\big[D_{\pi}(p+k)-D_\pi(p)\big]\nonumber\\
&-&\frac{2}{N}\frac{\delta \Delta V_{\pi}^0}{\delta G_{\pi}},
\eea
and we have shown in \cite{Fejos:2009dm} that the second integral is free of divergences. Exploiting \eqref{Eq:Gss} in the first (local) integral, we immediately see that this term splits into two divergent pieces:
\begin{subequations}
\label{Eq:piondiv}
\bea
\label{Eq:piondiv_a}
-\frac{1}{3N}\int_p \lambda(p) D_{\pi}(p)\bigg|_{\div}&=:& -\frac{\lambda}{3N}\tilde{T}_{\div}(M^2), \\
i\frac{v^2}{9N}\int_p \lambda^2(p)D^2_{\pi}(p)\bigg|_{\div}&=&-\frac{\lambda^2v^2}{9N}T_a^{(0)}.
\label{Eq:piondiv_b}
\eea
\end{subequations}
In \cite{Fejos:2009dm} we did not encounter a divergence proportional to $v^2$, as $T_a^{(0)}$ was considered finite, and furthermore, though the definition of $\tilde T_{\div}(M^2)$ remains the same, its expression changes. To obtain it, we need the expansions \eqref{Eq:prop_expand} and \eqref{Eq:eff_cpl_expand}, as well as the expansion of the difference $I_\pi^F(p)-I_0^F(p),$ which up to ${\cal O}(1/p^4)$ is given by
\bea
\label{Eq:I_expand}
I_\pi^F(p)-I_0^F(p)&\simeq&\frac{i}{8\pi^2}\left[3(M^2-M_0^2)-M^2\ln\frac{M^2}{M_0^2}\right] G_0(p)\nonumber\\
&+&2i(M^2-M_0^2)I_0^F(p) G_0(p)\,.
\eea
This can be derived using the explicit expression for $I_\pi^F(p)$ and $I_0^F(p)$ obtained by sending the cutoff to infinity in their defining integral. With a bit of algebra we obtain
\be
\tilde{T}_{\div}(M^2)=t_1(M^2)+\frac{\lambda}{3}T_a^{(0)}\int_p D_{\pi}(p),
\label{Eq:Tdiv}
\ee
where we introduced 
\begin{subequations}
\label{Eq:t12_of_M2}
\bea
\label{Eq:t2_of_M2}
t_1(M^2)&=&T_a^{(2)}-\frac{\lambda}{3}T_a^{(0)}T_d^{(2)} - (M^2-M_0^2) t_2,\\
t_2&=&\frac{\lambda}{2}T_a^{(I)}+T_a^{(0)}\left[\frac{\lambda}{3}\left(T_d^{(0)}+\frac{1}{8\pi^2}\right)-1\right],\qquad
\eea
\end{subequations}
and expressed $M^2\ln(M^2/M_0^2)$ in terms of the tadpole integral by using in \eqref{Eq:subt1_T} the explicit expression of the finite tadpole
\be
T_\pi^F=\frac{1}{16\pi^2}\left(M^2\ln\frac{M^2}{M_0^2}-M^2+M_0^2\right),
\label{Eq:TpiF}
\ee
again obtained for infinite cutoff.

With the help of (\ref{Eq:piondiv}) and (\ref{Eq:Tdiv}) the following expression for $\Delta V_{\pi}^0$ is determined from (\ref{Eq:pion}): 
\bea
\Delta V_{\pi}^0&=& -\frac{\lambda}{6}\left[t_1(M^2)+\frac{\lambda}{3} v^2 T_a^{(0)}\right]\int_p G_{\pi}(p)\nonumber\\
&&-\frac{\lambda^2}{36}T_a^{(0)}\left(\int_p G_{\pi}(p)\right)^2.
\label{Eq:deltaV_pi}
\eea
Note that $\Delta V_{\pi}^0$ depends linearly on $\hat{\alpha}$ through $M^2$, and that we got a new term proportional to $v^2$ and, furthermore, a double scoop integral (last term on the right-hand side). Terms of these types were not present in \cite{Fejos:2009dm} in the auxiliary field formulation of the model, however there is no symmetry restriction preventing its emergence in the counterterm functional. At NLO in the $1/N$ expansion $\Delta V_{\pi}^0$ gives the following finite pion propagator:
\bea
i G^{-1}_\pi(k)&=&k^2-M^2-\frac{\lambda}{3N}\bigg[i\int_p G_{\alpha\alpha}^{(0)}(p) D_\pi(k+p)\nonumber\\
&&-\tilde T_\textnormal{div}(M^2)-\frac{\lambda v^2}{3}T_a^{(0)}\bigg].
\label{Eq:pi-prop}   
\eea

\subsection{The derivative of the effective potential with respect to $v$} 

This derivative is given by
\bea
\frac{\delta V}{\delta v}=NvM^2-i\sqrt{\frac{\lambda}{3}}\int_k G_{\alpha \sigma}(k)+\frac{\delta \Delta V_{\pi}^0}{\delta v}+\frac{\delta \Delta V_{v}^0}{\delta v},\ \ \ \ 
\label{Eq:eos}
\eea
where we have indicated that there is also a contribution from the $\Delta V_{\pi}^0$ counterterm given in \eqref{Eq:deltaV_pi}. Upon using the LO expression of $G_{\alpha \sigma}$ from \eqref{Eq:Gas} together with \eqref{Eq:Gss} for the LO sigma propagator, one sees that the integral in \eqref{Eq:eos} splits into the same two divergent contributions given in \eqref{Eq:piondiv}, both appearing now with opposite sign. There is a dangerous environment dependent ({\it i.e.} temperature dependent in a finite temperature setting) subdivergence proportional to the tadpole $\int_k D_{\pi}(k)$  included in $\tilde{T}_{\div}$, but fortunately it exactly cancels with the term coming from $\delta \Delta V_{\pi}^0/\delta v$. The expression of $\Delta V_v^0$ is determined by the requirement of the cancellation of all remaining divergences in (\ref{Eq:eos}):
\bea
\label{Eq:deltaV_v}
\Delta V_v^0=-\frac{\lambda}{6} t_1(M^2)v^2-\frac{\lambda^2}{36}T_a^{(0)}v^4.
\eea
We see that a four-point counterterm vertex appeared as the last term on the right-hand side, a type of operator which was absent in \cite{Fejos:2009dm} in the auxiliary field formulation of the model, but as was the case with the new term emerging in the pion propagator equation, this is neither forbidden by any symmetry. We note that by construction Goldstone's theorem is respected with the present extended subtraction as well.

\subsection{The derivative of the effective potential with respect to $\hat{\alpha}$} This expression is given by
\bea
\frac{\delta V}{\delta \hat{\alpha}}&=&\frac{3N}{\lambda}\hat\alpha-i\frac{N}{2}\left(v^2+\int_k G_{\pi}(k)\right)+i\frac{N}{2}t_d(M^2)\nonumber\\
&-&\frac{i}{2}\int_k \big(G_{\sigma \sigma}(k)-G_{\pi}(k)\big)\nonumber\\
&+&\frac{\delta \Delta V_{\pi}^{0}}{\delta \hat{\alpha}}+\frac{\delta \Delta V_{v}^{0}}{\delta \hat{\alpha}}+\frac{\delta \Delta V_{\alpha}^{0}}{\delta \hat{\alpha}},
\label{Eq:spe}
\eea
where the term containing the expression $t_d(M^2)$ introduced in \eqref{Eq:td} is the contribution of $\Delta V_{\alpha}^N$ and renormalizes the expression at leading order. Note that, both $\Delta V_{\pi}^0$ and $\Delta V_{v}^0$ contribute to the right-hand side of (\ref{Eq:spe}):
\bea
\label{Eq:dVda}
\frac{\delta \Delta V_{\pi}^{0}}{\delta \hat{\alpha}}+\frac{\delta \Delta V_{v}^{0}}{\delta \hat{\alpha}}&=&-i\frac{\lambda}{6}t_2\left(v^2+\int_k D_{\pi}(k)\right),
\eea
where the term proportional to the tadpole comes from $\Delta V_{\pi}^0$, while the one containing $v^2$ arrives from $\Delta V_v^{0}$. The consistency of the procedure requires that $\Delta V_{\alpha}^0,$ the last piece of the counterterm functional left to be determined, depends on $\hat{\alpha}$ only, otherwise it would contribute to the pion propagator equation and/or to $\delta V/\delta v,$ and the procedure would not close. We also expect $\Delta V_{\alpha}^0$ to be a polynomial in $\hat{\alpha}$. 

There are two integrals in \eqref{Eq:spe} whose divergences have to be calculated. The first contains the difference of LO propagators at the needed accuracy in the $1/N$ expansion, and it is rather simple. Using \eqref{Eq:Gss} in \eqref{Eq:int_of_diff}, followed by expansions \eqref{Eq:prop_expand} and \eqref{Eq:eff_cpl_expand}, one obtains
\bea
\label{Eq:ss-pi_div}
\int_k \left[G^{(0)}_{\sigma\sigma}(k)-D_{\pi}(k)\right]\bigg|_{\div}
=\frac{\lambda v^2}{3} t_a.\ \ 
\eea
For the second integral we take the inverse of $G_\pi^{-1}$ given in \eqref{Eq:pi-prop} and expand it to ${\cal O}(1/N)$. Using the two integrals $J(M^2)$ and $\tilde J(M^2)$ introduced in Eq.~(29) of \cite{Fejos:2009dm}, and given also here for convenience
\begin{subequations}
\bea
\label{Eq:Js}
\tilde J(M^2)&=&\frac{1}{\lambda}\int_k D_\pi^2(k)\int_p \lambda(p) D_\pi(p+k),\ \ \ \ \ \\
J(M^2)&=&\frac{1}{\lambda^2}\int_k D_\pi^2(k) \int_p\lambda^2(p)D_\pi(p+k)G_{\sigma\sigma}^{(0)}(p), \ \ \ \ \ 
\eea
\end{subequations}   
one obtains
\bea
\int_k G_{\pi}(k)&=&\int_k D_\pi(k) 
-\frac{\lambda^2 v^2}{9N}\left[J(M^2)-i T_a^{(0)} \int_k D^2_\pi(k)\right]
\nonumber\\
&-&\frac{i\lambda}{3N}\left[\tilde J(M^2)- \tilde{T}_{\div}(M^2)\int_k D_\pi^2(k)\right].
\label{Eq:pi_nlo}
\eea
This replaces Eq.~(28) of \cite{Fejos:2009dm}, as it contains also the effect of the new subtraction. In order to isolate the divergences of $J$ and $\tilde J$ we change the order of integration, use the exact equality which holds at infinite cutoff,
\bea
\int_k D_{\pi}^2(k)D_{\pi}(p+k)&=&\frac{1}{p^2-4M^2}\bigg[I_\pi^F(p)+\frac{1}{8\pi^2}\nonumber\\
&&-\frac{1}{16\pi^2}\ln\frac{M^2}{M_0^2}\bigg],
\label{Eq:I2}
\eea
and expand the propagators around $G_0$ as in \eqref{Eq:prop_expand} and $\lambda(p)$ around $\lambda_0(p)$ using \eqref{Eq:eff_cpl_expand} and \eqref{Eq:I_expand}. A straightforward calculation yields
\begin{subequations}
\bea
\nonumber
\tilde{J}_{\div}(M^2)&=& - i\left[\frac{6}{\lambda}+T_d^{(0)}+\frac{1}{8\pi^2}\right]\left[\tilde T_{\div}(M^2)+3M^2t_a\right]\\\nonumber
&+&i\frac{6}{\lambda} t_d(4M^2) +3it_a M^2 I_\pi(k=0)\\
&+&\tilde T_{\div}(M^2)\int_k D_\pi^2(k)
\label{Eq:Jtilde}
\\
\label{Eq:J}
J_{\div}(M^2)&=& T_a^{(I)}+T_a^{(0)}\left(T_d^{(0)}+\frac{1}{8\pi^2}\right)\nonumber \\
&+&i T_a^{(0)}\int_k D^2_{\pi}(k),
\eea
\end{subequations}
where in both cases we replaced $\ln(M^2/M_0^2)$ by the finite  bubble integral at vanishing external momentum  obtained for infinite cutoff using the relation
\be
I_\pi^F(k=0) = \frac{1}{16\pi^2}\ln\frac{M^2}{M_0^2}, 
\label{Eq:IpiF}
\ee
and then used \eqref{Eq:subt1_I} to make appear the full bubble integral $I_\pi(0)$ at vanishing momentum.  

The last term of both (\ref{Eq:Jtilde}) and (\ref{Eq:J}) is a subdivergence which cancels immediately in (\ref{Eq:pi_nlo}). In order to make explicit another subdivergence of $\tilde J,$ related to the tadpole, we use \eqref{Eq:IpiF}, \eqref{Eq:subt1}, and \eqref{Eq:TpiF} to write
\bea
M^2 I_\pi(k=0)&=&\frac{M^2}{16\pi^2}\ln\frac{M^2}{M_0^2}+M^2 T_d^{(0)}\nonumber\\
&=&\int_k D_\pi(k) - t_d(0)+\frac{M^2-M_0^2}{16\pi^2}.
\eea
Using the above relation in the second line of \eqref{Eq:Jtilde} and \eqref{Eq:Tdiv} in the first line, one obtains the final expression:
\bea
\tilde J_{\div}(M^2)&=& - i\left[\frac{6}{\lambda}+T_d^{(0)}+\frac{1}{8\pi^2}\right]\left[t_1(M^2)+3M^2t_a\right]\nonumber\\
&+&i\frac{6}{\lambda} t_d(4M^2) -3it_a\left[t_d(0)+\frac{M_0^2-M^2}{16\pi^2}\right]\nonumber\\
&-&it_2\int_k D_\pi(k)+\tilde T_{\div}(M^2)\int_k D_\pi^2(k).
\label{Eq:Jtilde_final}
\eea

Among the contributions to \eqref{Eq:spe} there are dangerous terms proportional to $v^2$ and $\int_k D_{\pi}(k)$ which should disappear. The latter comes entirely from (\ref{Eq:Jtilde_final}) and (\ref{Eq:dVda}), which eventually cancel each other in (\ref{Eq:spe}). Concerning the terms proportional to $v^2$, first we combine (\ref{Eq:ss-pi_div}) with the corresponding term of (\ref{Eq:pi_nlo}) and realize that the result is exactly canceled by the remaining term of (\ref{Eq:dVda}). This means that there is no environment dependent subdivergence in (\ref{Eq:spe}) and the corresponding $\Delta V_{\alpha}^0$ counterterm depends only on a quadratic polynomial of $\hat{\alpha}$. Its final expression reads 
\be
\label{Eq:deltaV_a}
\Delta V_{\alpha}^0=i\hat\alpha \delta \kappa_1^{(1)}+\hat\alpha^2\delta\kappa_2^{(1)},
\ee
with
\bea
\delta\kappa_1^{(1)} &=& -\big(t_1(m^2)+3m^2t_a\big) \left[1+\frac{\lambda}{6}T_d^{(0)}+\frac{\lambda}{48\pi^2}\right]\nonumber\\
&+&t_d(4 m^2) -\frac{\lambda}{2}t_a\left(t_d(0)+\frac{M_0^2-m^2}{16\pi^2}\right),\\
\delta\kappa_2^{(1)} &=&\frac{t_2 - 3 t_a}{2}\left[1+\frac{\lambda}{6}T_d^{(0)}+\frac{\lambda}{48\pi^2}\right]+2 T_d^{(0)} +\frac{\lambda t_a}{64\pi^2},\nonumber\\
\eea
providing the two NLO order counterterms, exclusively related to the auxiliary field.

\subsection{The effective potential}

Putting together the different pieces of \eqref{Eq:cts} one recognizes that the new subtraction can be performed at NLO for arbitrary values of $v$ and $\hat\alpha$, in agreement with the general expectations on the structure of the counterterms. This completes the renormalization of the model in the auxiliary field formulation, where the counterterm functional is
\bea
\Delta V[\hat\alpha,v,G_\pi,{\cal G}]&=&
\frac{1}{2}\left(\delta \hat m^2-i\delta g\hat\alpha\right) \left(v^2+\int_k G_\pi(k)\right)
\nonumber\\ 
&+&i\delta\kappa_1\hat\alpha+\delta\kappa_2\hat\alpha^2
+ \frac{1}{2}\delta\kappa_0\int_k G_{\alpha\alpha}(k)
\nonumber\\ 
&+&\frac{\delta\hat\lambda}{4!} \left(v^2+\int_k G_{\pi}(k)\right)^2,
\label{Eq:Phi_ct-funct}
\eea
with the following countercouplings:
\bea
&&\delta g=\frac{\lambda}{3}t_2,\ \ \delta\hat\lambda=-\frac{2\lambda^2}{3}T_a^{(0)},\ \  \delta \hat m^2=-\frac{\lambda}{3}t_1(m^2),\nonumber \\ 
&&\delta \kappa_0=\frac{\lambda}{6}T_d^{(0)}, \ \ 
\delta \kappa_1=N\delta \kappa_1^{(0)}+\delta\kappa_1^{(1)},\ \  \delta \kappa_1^{(0)}=\frac{1}{2}t_d(m^2),\nonumber\\
&&\delta \kappa_2=N\delta \kappa_2^{(0)}+\delta \kappa_2^{(1)},\ \  \delta \kappa_2^{(0)}=\frac{1}{4}T_d^{(0)}.
\eea
The last term in \eqref{Eq:Phi_ct-funct} is a completely new functional term, compared to the expression in Eq.~(32) of \cite{Fejos:2009dm}, and one also notes that (\ref{Eq:Phi_ct-funct}) contains exclusively the sum of $v^2$ and $\int_kG_\pi(k).$ This feature is sufficient to preserve the validity of Goldstone's theorem also in the renormalized theory at NLO, since it ensures that the same subtraction is performed in both unrenormalized expressions of $\delta V/v\delta v$ and $2\delta V/\delta G_\pi,$ which is needed for the theorem to be obeyed. 
 
Combining \eqref{Eq:V_2PI} and \eqref{Eq:Phi_ct-funct}, we can define with $\delta\kappa_2$ the bare coupling $\lambda_B$ through the relation
\be
\frac{1}{\lambda_B}=\frac{1}{\lambda}+\frac{2}{3N} \delta\kappa_2.
\ee
Writing $\lambda_B=\lambda+\delta\lambda_\alpha$ and decomposing the counterterm into LO and NLO parts $\delta\lambda_\alpha=\delta\lambda_\alpha^{(0)}+\delta\lambda_\alpha^{(1)}/N$, the LO part of the coupling, $\lambda_B^{(0)}$ is determined by $\delta\kappa_2^{(0)}$ and reads
\be
\lambda_B^{(0)}=\frac{\lambda}{1+\lambda T_d^{(0)}/6},
\ee
and also
\be
\delta\lambda_\alpha^{(0)}=-\frac{\lambda^2}{6}\frac{\lambda}{1+\lambda T_d^{(0)}/6},\ \  
\delta\lambda_\alpha^{(1)}=-\frac{2}{3}\big(\lambda_B^{(0)}\big)^2\delta\kappa_2^{(1)}.
\ee

As we shall see in a moment, the LO part of the bare coupling remains unchanged even after the elimination of the auxiliary field, which we do in order obtain the ${\cal O}(N^0)$ accurate effective potential of the model as a functional of the original variables. Following Sec.~VI of \cite{Fejos:2009dm}, we need to substitute into \eqref{Eq:V_2PI} and \eqref{Eq:Phi_ct-funct} the LO expressions $G_{\alpha\alpha}^{(0)}$ and $G_{\alpha\sigma}^{(0)}$ expressed in terms of $G_{\sigma\sigma}^{(0)}\equiv G_{\sigma}$ and $G_{\pi}$, and to make use of the saddle point equation for $\hat{\alpha}.$ We do not present this procedure, as it was done in \cite{Fejos:2009dm} in quite some details. We only have to add the last term of \eqref{Eq:Phi_ct-funct} to the expression on the right-hand side of Eq.~(45) of \cite{Fejos:2009dm}. Defining the bare parameters of the model without the auxiliary field as
\bea
\lambda_b=\lambda_B\hat{c}^2+\frac{\delta\hat\lambda}{N},\quad m^2_b=m^2+\frac{\delta \hat m^2}{N}-\frac{\lambda_B\hat c\delta \kappa_1}{3N}\ , \ \ 
\eea
$(\hat c=1+\delta g/N)$ and using that to ${\cal O}(1/N)$
\be
\left[\lambda_B\hat c^2+\frac{\delta\hat\lambda}{N-1}\right]\frac{v^2}{12}\int_k G_\pi(k) 
\approx\frac{\lambda_b v^2}{12} \int_k G_\pi(k),
\ee
one obtains
\begin{widetext}
\bea
\label{Eq:V_2PI_final}
V[v,G_{\pi},G_{\sigma}]&=&\frac{N}{2}m_b^2v^2+\frac{N}{24}\lambda_b v^4-\frac{i}{2}\int_k (N-1)\left(\ln G_{\pi}^{-1}(k)+{\cal D}^{-1}_{\pi}(k)G_{\pi}(k)\right)
-\frac{i}{2}\int_k \left(\ln G_{\sigma}^{-1}+{\cal D}^{-1}_{\sigma}(k)G_{\sigma}(k)\right)\nonumber\\
&+&\frac{N}{24}\lambda_b\left(\int_k G_\pi(k)\right)^2+\frac{\lambda_B^{(0)}}{12}\int_k G_\pi(k)\int_p G_\sigma(p)
-\frac{\lambda_B^{(0)}}{12}\left(\int_k G_{\pi}(k)\right)^2 -\frac{i}{2}\int_k \ln\left(1-\frac{\lambda_B^{(0)}}{6}\Pi(k)\right)\nonumber\\
&-&\frac{\lambda_B^{(0)}}{6}v^2\int_k G_\sigma(k) + \frac{\lambda_B^{(0)}}{6}v^2\int_k \frac{G_\sigma(k)}{1-\lambda_B^{(0)}\Pi(k)/6},
\eea
\end{widetext}
where we have introduced the notation $\Pi(k)=-i\int_p G_{\pi}(p+k)G_{\pi}(p)$ and the tree-level propagators 
\begin{subequations}
\bea
i {\cal D}_\pi^{-1}(k)&=&k^2-m^2_b-\frac{\lambda_b}{6}v^2, \\
i {\cal D}_\sigma^{-1}(k)&=&k^2-{m^2_b}^{(0)}-\frac{\lambda_B^{(0)}}{2}v^2.
\eea
\end{subequations}
The interpretation of the last four terms in \eqref{Eq:V_2PI_final} in terms of Feynman diagrams was given in Eqs.(50) and (51) and Fig.~2 of \cite{Fejos:2009dm}. Note that in ${\cal D}_\sigma^{-1}$ we replaced $m^2_b$ by its leading order part ${m^2_b}^{(0)}$ because we are interested only in the ${\cal O}(N^0)$ accurate effective potential. As it might be expected, one has just a single bare squared mass $m_b^2$ and a single bare coupling $\lambda_b,$ but in some terms only the LO part of them, ${m^2_b}^{(0)}$ and $\lambda_B^{(0)}$ is needed. 

Using $\lambda_b=\lambda+\delta\lambda$ and $m_b^2=m^2+\delta m^2,$ as well as the decompositions $\delta\lambda =\delta\lambda^{(0)}+\delta\lambda^{(1)}/N$ and $\delta m^2 =\delta {m^2}^{(0)}+\delta {m^2}^{(1)}/N,$ one obtains the following LO and NLO countercouplings:
\bea
&&\delta {m^2}^{(0)}=-\frac{1}{3}\lambda_B^{(0)}\delta\kappa_1^{(0)},\nonumber\\
&&\delta {m^2}^{(1)}=\delta \hat m^2-\frac{1}{3}\big[\delta\lambda_\alpha^{(1)}\delta\kappa_1^{(0)}+\lambda_B^{(0)} \big(\delta\kappa_1^{(1)}+\delta\kappa_1^{(0)}\delta g\big) \big],\nonumber\\
&&\delta\lambda^{(0)}=\delta\lambda_\alpha^{(0)},\quad \delta\lambda^{(1)}=\delta\lambda_\alpha^{(1)}+ 2\lambda_B^{(0)}\delta g + \delta\hat\lambda,
\label{Eq:ct_no_aux_final}
\eea
where the correction represented by the last term in \eqref{Eq:Phi_ct-funct} shows up in the NLO coupling counterterm $\delta\lambda^{(1)}.$\\ 

\subsection{Renormalization without the auxiliary field}

The countercouplings given in \eqref{Eq:ct_no_aux_final} renormalize by their very construction the propagator equations for the pion and sigma fields derived from $V[v,G_\pi,G_\sigma],$ as well as the derivative $\delta V/\delta v.$ We mention that there is no need to use the auxiliary field method to obtain the expression of the countercouplings in the theory written in the original variables because we presented in \cite{Fejos:2009dm} a method to determine them starting from \eqref{Eq:V_2PI_final}.  When applied to the renormalization of the pion propagator, this method requires first to remove the divergence of the momentum-dependent part of the self-energy and then of the momentum-independent piece of it. The explicit expressions in the inverse of the pion propagator $i G_\pi^{-1}(k)=k^2-M^2-\lambda \Sigma_\pi^F(k)/(3N)$ are
\bea
&&\Sigma_\pi^F(k)=\int_p\left[\frac{1}{1-\lambda \Pi_F(p)/6}-\frac{\lambda v^2}{3} \frac{iG_\sigma(p)} {\left(1-\lambda \Pi_F(p)/6\right)^2} \right]\nonumber\\
&&\qquad\qquad\times G_\pi(k+p)-\tilde T_\textnormal{div}(M^2),
\label{Eq:non-local_SE}\\
&&M^2=m_b^2+\frac{\lambda_b}{6}\left[v^2+\int_k G_\pi(k)\right] +\frac{\lambda_b}{6 N}\int_k \big[G_\sigma(k)-G_\pi(k)\big]\nonumber\\ 
&&\qquad+\frac{\lambda}{3 N}\tilde T_\textnormal{div}(M^2),
\label{Eq:local_SE}
\eea
where the first line in the momentum-independent part $M^2$ is obtained from \eqref{Eq:V_2PI_final} by differentiating with respect to $G_\pi$ and the last term is added there to compensate for its subtraction from the momentum-dependent part, done to render it finite. Writing $M^2={M^2}^{(0)}+{M^2}^{(1)}/N$ and expanding $G_\pi$ to ${\cal O}(1/N),$ we obtain the same integrals which appear in the auxiliary field formulation of the model, but $D_\pi(p)$ originally defined below \eqref{Eq:I_def} has now $M^2$ replaced by ${M^2}^{(0)}.$ Referring to Eq.~(55) of \cite{Fejos:2009dm} for some details, below we only give the corrected equation from which the NLO countercouplings can be determined:
\bea
&&-\frac{3i}{\lambda_B^{(0)}}\left[\delta {m^2}^{(1)}+\frac{\lambda}{3}\tilde T_\textnormal{div}\big({M^2}^{(0)}\big)\right]\nonumber\\
&&=\frac{i\delta\lambda^{(1)}}{2\lambda_B^{(0)}}\left(v^2+\int_k D_\pi(k)\right)+\frac{i}{2}\int_k \Big[G_\sigma(k)-D_\pi(k)\Big]\bigg|_\textnormal{div}
\nonumber\\
&&\qquad +\frac{\lambda}{6}\left[\tilde J_\textnormal{div}({M^2}^{(0)})-\tilde T_\textnormal{div}({M^2}^{(0)})\int_k D_\pi^2(k) \right]\nonumber\\
&&\qquad -i\frac{\lambda^2}{18}v^2 \left[J_\textnormal{div}({M^2}^{(0)})-i T_a^{(0)}\int_k D_\pi^2(k)\right].
\label{NLO-countercouplings}
\eea
In order to obtain a relation which involves the countercouplings and ${M^2}^{(0)}$ we have to use the integrals \eqref{Eq:subt1_T}, \eqref{Eq:ss-pi_div}, \eqref{Eq:J}, and \eqref{Eq:Jtilde_final}. Then, we substitute in it the LO finite gap equation ${M^2}^{(0)}=m^2+\lambda(v^2+T_\pi^F)/6$ and determine $\delta\lambda^{(1)}$ by requiring the vanishing of the coefficient of $v^2+T_\pi^F$. The vanishing of the remainder in that relation determines $\delta {m^2}^{(1)}.$ With this procedure we arrive at the expressions given in \eqref{Eq:ct_no_aux_final}.

\section{Discussion and Conclusions \label{sec:concl}}

The present study shows that the subtraction of the new divergent integral omitted from the renormalization procedure discussed in Ref.~\cite{Fejos:2009dm} does not change one of its main conclusions, namely that the renormalization of the model in the auxiliary field formulation can be performed at arbitrary value of the auxiliary field. 

However, we could separate analytically the divergences of the encountered integrals only if in some finite integrals, {\it e.g.} the finite bubble $I_\pi^F$ defined below \eqref{Eq:I_def} and the integral in \eqref{Eq:I2}, the cutoff is sent to infinity. Since strictly speaking the presence of the Landau pole imposes a restriction on the maximal value of the momentum scale present in the effective theory, we should investigate how the divergence analysis goes in the case when this restriction is imposed on every subdiagram too. It turns out that in the auxiliary field formulation of the model, after the renormalization of the pion propagator and of the field equation,  the cancellation of subdivergences from the saddle point equation imposes a constraint among the above mentioned two integrals which has to be satisfied by a consistent cutoff regularization scheme. 

The emergence of the constraint is easily seen as follows. If $I_\pi^F$ and $I_0^F$ are calculated in Euclidean space with a yet unspecified cutoff regularization then on the right-hand side of the expression in \eqref{Eq:I_expand} an additional term $\Delta I^\Lambda(k_E;M^2,M_0^2)$ will appear due to the explicit dependence of the finite bubbles on the cutoff $\Lambda$. This term emerges from an expansion for small $M^2$ and $M_0^2$ without assuming $|k_E|\ll\Lambda$ and vanishes when $\Lambda\rightarrow\infty$. Then $t_1(M^2)$ in \eqref{Eq:Tdiv} gets a correction of the form $\frac{1}{6\lambda}\int_{k_E}^\Lambda\lambda_{0,E}^2(k_E) \Delta I^\Lambda(k_E;M^2,M_0^2) G_0(k_E),$ where  $\int_{k_E}^\Lambda = \int\frac{d^4 k_E}{(2\pi^2)}\theta(\Lambda-|k_E|),$ $G_0(k_E)=1/(k_E^2+M_0^2)$ and $\lambda_{0,E}(k_E)$ is the Euclidean continuation of $\lambda_0(k).$ Due to \eqref{Eq:deltaV_pi} and \eqref{Eq:deltaV_v} the derivative of this integral with respect to $\hat\alpha$ (or equivalently $i M^2$) will appear on the right-hand side of \eqref{Eq:dVda} as a $\Lambda$-dependent correction of the form $-\frac{i}{36}\int_{k_E}^\Lambda\lambda_{0,E}^2(k_E) G_0(k_E) \frac{d}{d\hat\alpha}\Delta I^\Lambda(k_E;M^2,M_0^2).$ Now, if the same regularization is applied to calculate the divergent part of $J_E(M^2)$ then the integral in \eqref{Eq:I2} (written as an Euclidean integral with Euclidean propagators) acquires on the right-hand side a $\Lambda$-dependent correction $\Delta I_2^\Lambda(k_E;M^2,M_0^2),$ so that the correction in $J_E(M^2)$ will be of the form $-\frac{1}{\lambda^2}\int_{k_E}^\Lambda\lambda_{0,E}^2(k_E) \Delta I_2^\Lambda(k_E;M^2,M_0^2) G_0(k_E).$ Then, as we can readily check, the cancellation of the $v^2$-dependent divergences in the saddle point equation \eqref{Eq:spe} occurs only if the relation
\be
\Delta I_2^\Lambda(k_E;M^2,M_0^2) = -\frac{1}{2} \frac{d}{dM^2} \Delta I^\Lambda(k_E;M^2,M_0^2),
\label{Eq:constraint}
\ee
is satisfied. The same relation is the precondition for the cancellation of the divergences proportional to $\int_{k_E} D(k_E)$. 

It is very plausible that the relation \eqref{Eq:constraint} holds for small $M^2$ and $M_0^2,$ because a similar one exists between $I_\pi^F$ and the integral in \eqref{Eq:I2} at infinite cutoff and also when one chooses at the level of the effective potential a regularization for which to every propagator a regulator function is attached (in case of a sharp cutoff the regulator is $\Theta(\Lambda - |q_E|)$ where $q_E$ stands for $k_E,p_E$ or $k_E+p_E$). Such a regularization of the 2PI effective action, which preserves the invariance of the unregularized, formal integral against shifts of the loop momenta which permute the arguments of the propagators, was discussed in \cite{Marko:2012wc}. In case of four-dimensional rotational invariant functions this regularization leads in the equation of $G_{\alpha\alpha}$ and $G_\pi$ to integrals of the form [see Eq.~(A1) of \cite{Fejos:2011zq}]
\bea
&&\int_{k_E}^\Lambda f(|k_E|) g(|k_E+p_E|)\theta(\Lambda-p) \theta(\Lambda-|k_E+p_E|)\nonumber\\
&&=\frac{\theta(\Lambda-p)}{8\pi^3p^2}\left(\int_0^{\Lambda-p} d k\, k f(k) \int_{|p-k|}^{k+p} d q\, q g(q) J(q)\right.\nonumber\\
&&\qquad\qquad\left.+\int_{\Lambda-p}^\Lambda d k\, k f(k) \int_{|p-k|}^\Lambda d q\, q g(q) J(q)\right),
\label{Eq:2theta_reg}
\eea
where $J(q)=[4k^2q^2-(q^2+k^2-p^2)]^{1/2}$ with $k=|k_E|$ and $p=|p_E|.$ In case of $f(k)=1/(k^2+M^2)^n$ with $n=1$ or $n=2$ and $g(k)=1/(k^2+M^2)$ the second double integral cannot be calculated analytically, making difficult to obtain explicitly the corrections $\Delta I^\Lambda(k_E;M^2,M_0^2)$ and $\Delta I_2^\Lambda(k_E;M^2,M_0^2).$ 

If instead of attaching a regulator to each propagator, one cuts only the loop momenta in the setting-sun diagram of \eqref{Eq:V_2PI} one finds in the equation for $G_{\alpha\alpha}$ a bubble integral similar to \eqref{Eq:2theta_reg} but without the second theta function $\theta(\Lambda-|k_E+p_E|)$. In this case the second double integral is missing from \eqref{Eq:2theta_reg} and the upper limit of the outer integral in the first double integral is $\Lambda$ instead of $\Lambda-p.$ In this regularization $\Delta I^\Lambda(k_E;M^2,M_0^2)$ can be calculated explicitly by a direct calculation. The expression of $\Delta I_2^\Lambda(k_E;M^2,M_0^2),$ obtained indirectly from the consistency relation \eqref{Eq:constraint} determines $J_E^{\div}(M^2)$. However, as one can check numerically, this does not render finite the integral $J_E(M^2),$ which, in the regularization when only the loop momenta in the setting-sun diagram are cut, turns out to be written in terms of the average of the integral in \eqref{Eq:I2}, calculated with the one-theta and the two-theta regularizations discussed above. This shows that cutting only the loop momenta is not a consistent regularization. After all, this should not come as a surprise because even the starting expression of the regularized setting-sun diagram in the effective potential changes if we cut the loop momenta after shifting them (permuting the arguments of the propagators). Furthermore, if we remind ourselves that the two-loop setting-sun diagram is originally obtained by integrating over all momenta of its three propagators in the presence of a delta function ensuring momentum conservation, for consistency reasons actually all three integrals in question should be cut. Performing one of them with the help of the delta function however leads to exactly the same regularization discussed above and in \cite{Marko:2012wc}.

\begin{figure}[!t]
\begin{center}
\includegraphics[width=0.49\textwidth,angle=0]{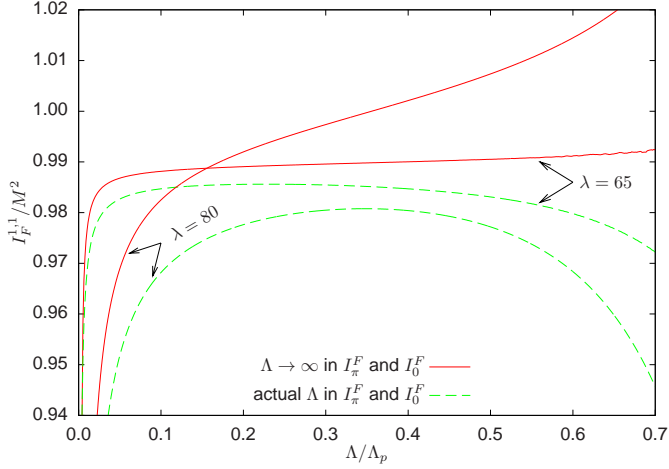}
\caption{The cutoff dependence of the subtracted integral $I_F^{1,1}$ for two ways of computing the finite bubble integrals $I_\pi^F(p)$ and $I_0^F(p)$: with the actual value of the cutoff or by taking $\Lambda\to\infty.$ The mass parameters are those of Fig.~\ref{Fig:Inm} and for the sake of presentation the two curves with $\lambda=80$ were shifted downwards by 0.385.}
\label{Fig:I11}
\end{center}
\end{figure}

There is no such consistency problem in the formulation of the model not using the auxiliary field. In this case one can regularize the integrals of the pion propagator equation appearing in \eqref{Eq:non-local_SE} and \eqref{Eq:local_SE} by cutting the loop integrals only. In this case the finite bubble integrals $I_\pi^F(p)$ and $I_0^F(p),$ and the integral in \eqref{Eq:I2} can be evaluated with a finite cutoff and corrections to counterterms, arising from keeping the cutoff in the expression of these integrals, can be calculated. For example, with an explicit calculation one obtains that the correction $\Delta I^\Lambda(k_E;M^2,M_0^2)$ is proportional to $M^2-M_0^2$ such that $t_2$ changes to 
\be
t_2^\Lambda = t_2 + \frac{\lambda}{96\pi^2}\left(\frac{d^{(2)}}{\Lambda^2}+L_a^{(0)}\right),
\ee
with $d^{(2)}=(1/\lambda^2)\int_{p_E}^\Lambda G_0(k_E)\lambda_{0,E}^2(p_E)$ and $L_a^{(2)}=(1/\lambda^2)\int_{p_E}^\Lambda G_0^2(k_E)\lambda_{0,E}^2(p_E)\ln(1-p_E^2/\Lambda^2).$ Note that we have kept those integrals from which constant contributions would arise in $t_2^\Lambda$ for $\Lambda\to\infty$ in the absence of $\lambda_{0,E}^2(p_E)$ from their integrand. Using the above relation in the expression of $\tilde T_{\div}(M^2)$ one can compare the cutoff dependence of the finite integral $I_F^{1,1}=I^{1,1}-\tilde T_{\div}(M^2)$ when the finite bubbles are calculated with infinite or finite cutoff. This is presented in Fig.~\ref{Fig:I11} for two different values of the coupling. For the smaller coupling a plateau-like behavior can be seen in both cases. This behavior is even more pronounced for smaller values of the cutoff, in which case the two curves are closer to each other in the region where one could speak about an apparent convergence of the integral with the increasing of $\Lambda.$ The $\Lambda$ dependence of $I_F^{1,1}$ is different at large values of the cutoff: it has an inflection when the finite bubbles are calculated with an infinite cutoff and a maximum when they are calculated with the actual value of the cutoff.

\medskip

All $n$-point functions derived from (\ref{Eq:V_2PI_final}) with the appropriately chosen countercouplings are consistently freed from all divergent (i.e. strongly cutoff dependent) contributions. Still, all "convergent'' NLO contributions to these observables of the theory are sensitive to the Landau pole through the presence of $\lambda(p)$ in the integrands of the contributing integrals $I^{j,k},~j>2$. One might contemplate to apply further subtractions with the aim to decrease the range of influence of the Landau pole in specific $n$-point functions around its actual location. As an example, one can use in the renormalized equations of $G_\pi$ and $v$ the iterated version of \eqref{Eq:Gss} in \eqref{Eq:Gaa} and keep more terms in the expansion of $D_\pi(p)$ and $\lambda(p)$ around $G_0(p)$ and $\lambda_0(p),$ respectively. (One has to synchronize this oversubtraction between the two equations in order to maintain Goldstone's theorem.) For instance, one might define the finite part of the integral appearing on the left-hand side of \eqref{Eq:piondiv_b} by subtracting also the first convergent terms in the expansion of its integrand around $\lambda_0^2(p) G^2_0(p).$ Using \eqref{Eq:I_expand} and neglecting for simplicity $T_\pi^F$ obtained by rewriting $M^2 \ln(M^2/M_0^2)$ with the help of \eqref{Eq:TpiF}, we obtain 
\be
\tilde I_F^{2,2}:=I^{2,2}-i\lambda^2\tilde T_a^{(0)},
\ee
where $\tilde T_a^{(0)}=T_a^{(0)}+(M^2-M_0^2)F(M_0)/\lambda^2,$ with $F(M_0)=\big[6 I_0^{3,2}-4 (\lambda^{-1}+1/(48\pi^2))I_0^{3,3}\big]$ given in terms of $I_0^{j,k},$ the integral defined in \eqref{Eq:I_def}, but with $M^2$ replaced by $M_0^2.$ Then, compared to \eqref{Eq:piondiv_a}, we can choose to define a modified finite part of $I^{1,1}$ using the replacement $T_a^{(0)}\to \tilde T_a^{(0)}$ in \eqref{Eq:Tdiv}, namely: 
\be
\tilde I_F^{1,1}:=I^{1,1}-\lambda\left(t_1(M^2)+\frac{\lambda}{3}\tilde T_a^{(0)}\int_k D_\pi(k)\right).
\ee
With this choice of oversubtraction the algebraic structure of the divergence cancellation does not change neither in the NLO pion propagator nor in $\delta V/\delta v.$ We only have to perform the change $T_a^{(0)}\to\tilde T_a^{(0)}$ in \eqref{Eq:deltaV_pi} which in turn induces the same change in \eqref{Eq:deltaV_v} and these two together lead in place of (\ref{Eq:dVda}) to
\bea
\frac{\delta\Delta V_\pi^0}{\delta \hat\alpha}+\frac{\delta\Delta V_v^0}{\delta \hat\alpha}&=&-i\frac{\lambda}{6}t_2\left(v^2+\int_k D_\pi(k)\right)\nonumber\\
&-&i\frac{\lambda}{36}F(M_0)\left(v^2+\int_k D_\pi(k)\right)^2\,.\ \ \ 
\eea
One should investigate if the second term on the right-hand side is canceled by the integrals defining the saddle point equation of $\hat{\alpha}$. It might not be possible to fulfill this {\it ad hoc} requirement for general values of $\hat\alpha$, but even then one easily constructs an appropriately defined term for $\Delta V_\alpha$ which would cancel it at the specific $\hat\alpha$ value satisfying the LO saddle point equation. Even without this compensation this term represents just a finite(!) contribution to the saddle point equation. An appropriate choice of $F(M_0)$ might diminish the effect of the Landau pole in the field and pion propagator equations, while other $n$-point functions might receive extra Landau-pole sensitive contributions due to the extra counterterms produced by the oversubtraction.

\medskip

In conclusions, we revisited the problem of renormalizing the $O(N)$ model at NLO in the $1/N$ expansion. This was necessary, because, although we were aware of the presence of the Landau pole, the renormalization performed in \cite{Fejos:2009dm} was based on the behavior of the integrands at asymptotically large momenta. Now we focused our discussion on defining a cutoff insensitive effective potential with a cutoff below the scale of the Landau singularity. It turned out that more care is needed in the study of the divergences because in some cases the behavior of the integrand is different below and above the singularity. As a result of the subtraction of the integral $T_a^{(0)}$ defined in \eqref{Eq:Ta0} a rather important improvement of the cutoff insensitivity was experienced already below the Landau pole. 

\begin{acknowledgments}
G.~F. and Zs.~Sz. would like to thank Urko Reinosa for clarifying discussions leading to the conclusion that the integral $T_a^{(0)}$ defined in \eqref{Eq:Ta0} needs to be subtracted and the renormalization procedure of \cite{Fejos:2009dm} reanalyzed, as well as for various exchanges of ideas on the meaning of the renormalization in the presence of a Landau pole and the classification of integrals into convergent and divergent ones in that case. A.~P. and Zs.~Sz. were supported by the Hungarian Research Fund (OTKA) under Contract No. K104292. G.~F. is supported by the Foreign Postdoctoral Program of RIKEN.
\end{acknowledgments}


\begin{thebibliography}{99}
\bibitem{Root:1974zr} 
  R. G.~Root,
  Phys.\ Rev.\ D {\bf 10}, 3322 (1974).

\bibitem{Andersen:2008qk} 
  J.~O.~Andersen and T.~Brauner,
  Phys.\ Rev.\ D {\bf 78}, 014030 (2008).

\bibitem{Fejos:2009dm} 
  G.~Fej\H{o}s, A.~Patk{\'o}s, and Zs.~Sz{\'e}p,
  Phys.\ Rev.\ D {\bf 80}, 025015 (2009).

\bibitem{Cooper:2005vw} 
  F.~Cooper, J. F.~Dawson, and B.~Mihaila,
  Phys.\ Rev.\ D {\bf 71}, 096003 (2005).

\bibitem{Berges:2005hc} 
  J.~Berges, Sz.~Bors\'anyi, U.~Reinosa, and J.~Serreau,
  Ann. Phys. (Amsterdam)  {\bf 320}, 344 (2005).

\bibitem{Schnitzer:1974ji} 
  H.~J.~Schnitzer,
  Phys.\ Rev.\ D {\bf 10}, 1800 (1974).

\bibitem{Coleman:1974jh} 
  S. R.~Coleman, R.~Jackiw, and H. D.~Politzer,
  Phys.\ Rev.\ D {\bf 10}, 2491 (1974).

\bibitem{Abbott:1975bn} 
  L. F.~Abbott, J.~S.~Kang, and H. J.~Schnitzer,
  Phys.\ Rev.\ D {\bf 13}, 2212 (1976).

\bibitem{Bardeen:1983st} 
  W. A.~Bardeen and M.~Moshe,
  Phys.\ Rev.\ D {\bf 28}, 1372 (1983).

\bibitem{Nunes:1993bk} 
  J. P.~Nunes and H. J.~Schnitzer,
  Int.\ J.\ Mod.\ Phys.\ A {\bf 10}, 719 (1995).

\bibitem{Reinosa:2011ut} 
  U.~Reinosa and Zs.~Sz{\'e}p,
  Phys.\ Rev.\ D {\bf 83}, 125026 (2011).

\bibitem{Marko:2012wc} 
  G.~Mark{\'o}, U.~Reinosa, and Zs.~Sz{\'e}p,
  Phys.\ Rev.\ D {\bf 86}, 085031 (2012).

\bibitem{Marko:2013lxa} 
  G.~Mark{\'o}, U.~Reinosa, and Zs.~Sz{\'e}p,
  Phys.\ Rev.\ D {\bf 87}, 105001 (2013).

\bibitem{Mihaila:2000sr} 
  B.~Mihaila, F.~Cooper, and J. F.~Dawson,
  Phys.\ Rev.\ D {\bf 63}, 096003 (2001).

\bibitem{Aarts:2002dj} 
  G.~Aarts, D.~Ahrensmeier, R.~Baier, J.~Berges, and J.~Serreau,
  Phys.\ Rev.\ D {\bf 66}, 045008 (2002).

\bibitem{Fejos:2011zq} 
  G.~Fej\H{o}s and Zs.~Sz{\'e}p,
  Phys.\ Rev.\ D {\bf 84}, 056001 (2011).

\end{thebibliography}
\end{document}